\documentclass[preprint,preprintnumbers,amsmath,amssymb]{revtex4-2}
\usepackage{amsfonts}    
\usepackage{amssymb}
\usepackage{latexsym}
\usepackage{graphicx}
\usepackage{verbatim}
\usepackage{rotating}
\usepackage{multirow}
\usepackage[usenames,dvipsnames]{color}
\usepackage[active]{srcltx}
\usepackage[colorlinks=true]{hyperref}
\usepackage{color,soul}

\newcommand{\al}{\alpha}
\newcommand{\pa}{\partial}

\newcommand{\veps}{\varepsilon}

\newcommand{\la}{\lambda}

\newcommand{\Om}{\Omega}

\newcommand{\De}{\Delta}

\newcommand{\tha}{\theta}

\newcommand{\rar}{\rightarrow}
\newcommand{\lrar}{\leftrightarrow}
\newcommand{\re}[1]{(\ref{#1})}

\newcommand{\non}{\nonumber}
\begin{document}

\title{Helium-like ions $(Z,e,e)$ in Lagrange Mesh method, interpolating the highly-accurate energy spectra {\it vs.} $Z$}

\author{Horacio~Olivares-Pil\'on}
\email{horop@xanum.uam.mx}
\affiliation{Departamento de F\'isica, Universidad Aut\'onoma Metropolitana-Iztapalapa,
Apartado Postal 55-534, 09340 M\'exico, D.F., Mexico}

\author{Juan Carlos Lopez Vieyra}

\email{vieyra@nucleares.unam.mx}

\author{Alexander~V.~Turbiner}
\email{turbiner@nucleares.unam.mx, alexander.turbiner@stonybrook.edu}
\affiliation{Instituto de Ciencias Nucleares, Universidad Nacional
Aut\'onoma de M\'exico, Apartado Postal 70-543, 04510 M\'exico,
D.F., Mexico{}}

\begin{abstract}
Two alternative approaches for studying Helium-like atomic ions in non-relativistic quantum mechanics 
are proposed: 
(I) a numerical approach, based on the Lagrange-mesh method which can easily reach 
up to 14-15 significant digits in the energy spectrum for any nuclear charge $Z$ with modest 
CPU time in single processor mode and (II) a highly-accurate, few-parametric interpolation 
formula for the energies {\it vs.} $Z$.
The interpolation formula of general nature is proposed, it can be applied to the energies 
of any excited state of the helium-like sequence. It is based on matching the $1/Z$-expansion at large $Z$ 
and the Puiseux expansion with integer and half-integer powers around the so-called 
second critical charge $Z_B$, introduced by F and D Stillinger (1969, 1974), confirmed 
by the present authors in 2019 for the ground state $1^1 S$, then revisited here, and extended 
to the excited states in the present work. For example for the first two spin-singlet $1^1 S$, $2^1 S$ 
and the first two spin-triplet $2^3 S$, $3^3 S$ states this interpolation formula 
with nine free parameters can reach an accuracy of 10-14 significant digits (s.d.) in the energies 
for any physically-relevant nuclear charge $Z$, giving absolute accuracy at large $Z$.
Many results are obtained for the first time.
\end{abstract}

\date{8/9/26}

\maketitle
\section*{Introduction}

From the inception of the Schr\"odinger equation in 1926, the three-body quantum Coulomb problem has always 
attracted a lot of attention, especially, its one-center case where one Coulomb charge is assumed to be
infinitely massive. The Helium atom and helium-like sequence (equivalently, two-electron atomic ions) $(Z;e,e)$ 
are the most notable examples. From the time of Slater and Hylleraas, see e.g. \cite{Hylleraas}, 
the atomic physics community spent enormous efforts to calculate the energies of the low-lying states 
of the helium-like sequence: it can be easily found in the literature that there exist hundreds of papers 
where such calculations were carried out. Recently, the culmination was reached: Nakashima-Nakatsuji 
in 2007 \cite{Nakashima:2007} by using a variational trial function with $\gtrsim 20,000$ terms found 
the ground state energy for $Z=1 - 10$ 
with 41 significant digits (s.d.) which were mainly confirmed in 2018 \cite{Korobov:2018} for $Z=1,2$, 
again in another variational calculation but with a different trial function containing $\gtrsim 20,000$ terms. 
In 2008 Nakashima-Nakatsuji \cite{Nakashima:2008} extended their calculations to several low-lying excited states, 
where again unprecedented accuracies of about 40 s.d. were reached for the energy spectra.  
Similar highly accurate calculations for the low-lying excited states were also carried out in \cite{YP:2010} 
for $Z=3, \dots,12$ and also in \cite{Korobov:2018} for $Z=2$. In general, the results obtained in 
\cite{Nakashima:2007,Nakashima:2008} were confirmed for a large number of figures 
\footnote{Needless to say that such accuracies are well beyond of today's physics reach, both experimentally 
and theoretically. However, they can be useful in search for a new physics and also in mathematics studies 
of three-body Coulomb systems.}. A
natural question occurs about the ground state energies for larger $Z > 10$ and 
the energies of the low-lying and Rydberg excited states for $Z > 10$: should such 
enormous (and expensive) variational calculations be repeated for each larger value of $Z$? 
This question was addressed in \cite{OT:2015}, where the ground state energy of the negative Hydrogen ion $H^-$ 
was calculated numerically using the Lagrange mesh method (see for discussion \cite{HB:2001,Baye:2015})  
with 14 decimal digits (d.d.) and then in \cite{AoP:2019}, where the Lagrange mesh method was used 
to obtain 13-14 correct s.d. for the ground state energy of two-electron atomic ions with various 
$Z \leq 50$. These calculations, which are an alternative to variational studies, 
were very fast and much more economical, and they were in agreement with variational results 
wherever the comparison can be made.

In \cite{AoP:2019} a simple analytic interpolation with a few fitting parameters was proposed, 
which reproduced 13-14 s.d. for the ground state energy of all two-electron 
ions in the static approximation for different nuclear charges $Z \leq 50$. 
The goal of this article is twofold: (i) to demonstrate that the Lagrange Mesh Method 
provides a highly efficient numerical alternative to variational studies allowing 
to obtain up to 14-15 s.d. in the energies and (ii) to present the highly-accurate few parametric 
approximate expression for the energy of {\it any} excited state for the helium-like sequence
valid for all physically relevant nuclear charges. 
The idea behind this approximation is to {\it match} the celebrated $1/Z$-expansion at large $Z$, 
introduced by Hylleraas \cite{Hylleraas} in the early days of quantum mechanics, and the Puiseux expansion 
(in fractional degrees) around the so-called second critical charge $Z = Z_B$, where the bound state 
in question ceases to exist, as hinted long ago by F and D Stillinger \cite{Stillinger:1966} 
(and developed in \cite{AoP:2019}), into a {\it single function} in the form of the ratio of two polynomials 
in the variable $\sqrt {Z-Z_B}$. Note that both expansions exist and can be explicitly constructed 
for any excited state.
As an example, the two spin-singlet states, $1^1\,S$ (ground state) and $2^1\,S$, 
and the two spin-triplet lowest energy states, $2^3\,S$ and $3^3\,S$, will be considered.

The non-relativistic Coulomb system of $2$ electrons and an infinitely-heavy, static, 
point-like nuclear charge $Z$: $(Z; e, e)$, is described by the Hamiltonian
\begin{equation}
\label{CalH}
  {\cal H}\ =\ -\frac{1}{2\,m} (\De_1 + \De_2) \ -\ \frac{Z}{r_1} \ -\ \frac{Z}{r_2} \ +\
  \frac{1}{r_{12}}\ ,
\end{equation}
where $r_i$ is the distance from charge Z to $i$th electron (of mass $m=1$ and
charge $e=-1$), $\De_i$ is three-dimensional Laplacian associated with $i$th
electron, $r_{12}$ is the distance between electrons, and $\hbar=1$. The
scale transformation by Hylleraas: $r \rar r/Z$ \cite{Hylleraas}, leads to,
\begin{equation}
\label{H}
  {\cal H}\ =\ Z^2 H\  
  \equiv \ Z^2 \left(-\frac{1}{2} (\De_1 + \De_2) \ -\ \frac{1}{r_1} \ -\ \frac{1}{r_2} \ +\
  \frac{1}{Z}\, \frac{1}{r_{12}} \right) \ ,
\end{equation}
where $H$ is the Hamiltonian of negative hydrogenic ion $H^-$ with interelectron interaction 
of strength $1/Z$. In the limit $Z \rar \infty$ the Hamiltonian $H$ corresponds to 
two non-interacting Hydrogen atoms with analytically-known eigenvalues,
\begin{equation}
\label{E2H}
   E_{n_1,n_2}\ =\ -\frac{Z^2}{2}\,\left(\frac{1}{n_1^2} + \frac{1}{n_2^2}\right)\ ,
   \quad n_1/n_2\ =\ 1,2,3,\ldots\ ,
\end{equation}
where $n_1 (n_2)$ is the principal quantum number for the first (second) Hydrogen atom. 
It must be emphasized that for the $S$-states the eigenfunctions of ${\cal H}$ depend 
on the relative distances $r_1, r_2, r_{12}$ only and the Schr\"odinger equation becomes three-dimensional. 
This equation describes one particle in three-dimensional curved space \cite{TME:2017}. 

Throughout the paper all energies are given in atomic units (a.u.).
We use the abbreviation {\it s.d.} for {\it significant digits} and {\it d.d.}
for {\it decimal digits} throughout the text.

\section{\bf Solving the Schr\"odinger equation for $S$-states using the Lagrange Mesh Method (LMM).}

The Schr\"odinger equation is
\[
    {\cal H}\ \Psi\ =\ E\ \Psi \ ,\ \mbox{or}\ ,\ {H}\ \Psi\ =\ \veps\ \Psi
\]
with the Hamiltonian ${\cal H}/H$ (\ref{H}) defined in six-dimensional space 
${\bf R^6}={\bf R^3}(\vec{r_1})\oplus{\bf R^3}(\vec{r_2})$. 
Evidently, vectors $\vec{r_1},\vec{r_2},\vec{r_{12}}$ form triangle. Let us introduce the new variables 
$(r_1,r_2,r_{12},\Om)$, where $r$'s are relative distances and $\Om=(\tha_1,\tha_2,\tha_3)$ is 
a three-dimensional angle, which we will not define explicitly. As a result 
the six-dimensional Laplacian (the electronic kinetic energy) takes the form,
\[
   \De_1 + \De_2\ =\ \frac{1}{2}  \De_r (\rho_1=r_1^2,\rho_2=r_2^2,\rho_{12}=r_{12}^2)\ +\ 
   \De_{\Om}(r_1,r_2,r_{12},\Om; \pa_{\Om})\ ,
\]
see \cite{TME:2017}. Here 
\begin{equation}
\label{De_r}
  \De_r\ =\ 2 \rho_{1}\, \pa_{\rho_{1}}^2 +
  4 \rho_{12}\, \pa_{\rho_{12}}^2 +
  2 \rho_{2}\,\pa_{\rho_{2}}^2 +
\end{equation}
\[
  2(\rho_{1} + \rho_{12} - \rho_{2})\pa_{\rho_{1}\rho_{12}} +
  2(\rho_{2} + \rho_{12} - \rho_{1})\pa_{\rho_{2}\rho_{12}} +
  3 (\pa_{\rho_{1}} + 2 \pa_{\rho_{12}} + \pa_{\rho_{2}})\ ,
\]
and $\De_{\Om}$ is proportional to the derivatives with respect to angles {\it only}, hence, it annihilates any angle-independent function. If the potential does not contain any angle dependence, which is our case, 
this implies the existence of angle-independent eigenfunctions, which correspond to the $S$-states. 
Hence, for the $S$ states, the six-dimensional Schr\"odinger equation is reduced to the three-dimensional one,
\begin{equation}
\label{De_r-SE} 
    \bigg(- \frac{1}{2}\De_r \ -\ \frac{Z}{\sqrt \rho_1} \ -\ \frac{Z}{\sqrt \rho_2} \ +\ \frac{1}{\sqrt \rho_{12}} \bigg)\,\Psi (\rho_1,\rho_2,\rho_{12})\ =\ E\,\Psi (\rho_1,\rho_2,\rho_{12})\ ,
\end{equation}
with domain $0 \leq \rho_1,\rho_2,\rho_{12} \leq \infty$ and $\rho_{12} \leq \rho_1 + \rho_2$. 
Namely, this equation might be used for the LMM calculations. However, for practical 
reasons~\cite{HB:2001,Baye:2015} the calculations by using the Lagrange mesh method 
were carried out by solving equation (\ref{De_r-SE}) in perimetric coordinates, 
see below and also \cite{OT:2015,AoP:2019}. 
Note that equation (\ref{De_r-SE}) is the permutation-invariant, hence, 
it admits two types of solutions: 
symmetric, 
$$\Psi (\rho_2,\rho_1,\rho_{12})=\Psi (\rho_1,\rho_2,\rho_{12})\ ,$$ 
and anti-symmetric 
$$\Psi (\rho_2,\rho_1,\rho_{12})=-\Psi (\rho_1,\rho_2,\rho_{12})\ ,$$
with respect to permutation of electrons $1 \lrar 2$. These solutions correspond 
to spin-singlet (ortho-helium-like) or 
spin-triplet (para-helium-like) states, respectively.

\begin{figure}[h!]
\includegraphics[scale=2.0]{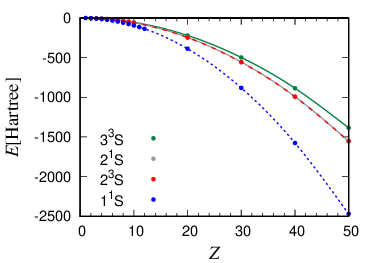}
\caption{Energy of the helium-isoelectronic sequence as a function of the  charge $Z$  
for the four states: ($i$) the ground state $1^1S$ (blue dotted line),
($ii$) the first spin-triplet state $2^3S$ (red dashed line), which is the next state after
the ground state and which is very close to the spin-singlet excited state $2^1S$ 
(gray dashed line) but it lies slightly below and ($iii$) the spin-triplet excited state $3^3S$ 
(solid green line).}
\label{fHel4S}
\end{figure}

The results of the LMM calculations for the energies of the two lowest spin-singlet states 
$1^1\,S$ (ground state) and $2^1\,S$, and the two lowest spin-triplet states, $2^3\,S$ and $3^3\,S$ 
for $Z \leq 50$ are presented in Tables \ref{EnSS21}-\ref{tableE33S}, 
respectively (see also Figure~\ref{fHel4S} as illustration). Typically, a $50 \times 50 \times 40$ lattice
in perimetric coordinates of the form,
\[
x= r_1- r_2 +r_{12}\ ,
\]
\[
y=-r_1+r_2+r_{12} \ ,
\]
\[
z=r_1+r_2-r_{12} \ ,
 \]
is used for calculations, it can reach an accuracy of 14 - 15 s.d. The details of the calculations
are presented in \cite{HB:1999, AoP:2019} for the case of the ground state $1^1\,S$: in the present work 
by using the same computer code the energies of the excited states are calculated. A standard laptop 
with 2.4 GHz single core processor can be sufficient to carry out the calculation with modest CPU times.  

\begin{table} 
\caption{Ground state $1^1S$ of the Helium  isoelectronic sequence for selected values of $Z$,
from $Z=1$ to $Z=50$.
In 2nd column the energies $E$ obtained via the LMM are shown: they are in agreement with those 
reported in~\cite{Nakashima:2007} (rounded), see also \cite{AoP:2019}: 
{\it all 14-15 printed digits exact}.
For $Z=20-50$ (shown in bold), there are no accurate results to compare with - 
this is the first accurate calculation.
The energies $E$ obtained via interpolation (13) shown in 3rd column 
(no rounding): {\it 12-14 digits exact}. 
At $Z \rar \infty$, the exact energy $E=E_{1,1}=-Z^2$.}
\scalebox{1.0}{
\begin{tabular}{c|c|c}
\hline
$Z$\ &\ Present (LMM)\ &\ $E_{\{2,2\}}^{(1^1S)}$~\re{fitP95}  \\
\hline
1 &\ -0.527751016544377\ &\ -0.52775101654437 \\
2 & -2.90372437703412 & -2.9037243770342 \\
3 & -7.27991341266931 & -7.2799134126682 \\
4 & -13.6555662384236 & -13.655566238430 \\
5 & -22.0309715802428 & -22.030971580234 \\
6 & -32.4062466018985 & -32.406246601894 \\
7 & -44.7814451487727 & -44.781445148773 \\
8 & -59.1565951227579 & -59.156595122763 \\
9 & -75.5317123639595 & -75.531712363970 \\
10& -93.9068065150375 & -93.906806515054  \\
20& {\bf -387.6572338332}   & -387.65723383319 \\
30& {\bf -881.4073774883}   & -881.40737748835  \\
40& {\bf -1575.1574495256}  & -1575.1574495255 \\
50& {\bf -2468.9074928127}  & -2468.9074928127 \\
\hline\hline
\end{tabular}}
\label{EnSS21}
\end{table}

\begin{table}
\caption{Energy $E$ of the excited state $2^1S$ for the Helium isoelectronic sequence. The
2nd column contains the previously reported results from [4-5, 13-15]. The energies 
obtained via the LMM ({\it all 14 printed digits exact}) and by using the interpolation shown in the 3rd 
and 4th columns, respectively.
For $Z=20-50$ (shown in bold), there are no accurate results to compare with - this is {\it the first calculation}.
At $Z \rar \infty$, the exact energy $E=E_{1,2}=-\frac{5}{8} Z^2$
\footnote{The accuracies of the results presented in [15] are grossly exaggerated. Their real accuracy is 6-8 s.d. for $Z \geq 20$. } 
.}
\scalebox{1}{
\begin{tabular}{c| l | c | c }
\hline
\hline
$Z$\ &\ \hspace{1.5cm}Refs [5, 11-14]\ &\ Present (LMM)\ &\  $E_{\{2,2\}}^{(2^1S)}$~\re{fitP95} \\
\hline
2  &\ -2.14597404605428~\cite{GD:1988H} \ &\ -2.1459740460544\ &\ -2.1459740460546 \\
   &\ -2.145974046054419~\cite{GD:2006}     & &                 \\
   &\ -2.1459740460544174~\cite{Korobov:2018}  & &                 \\
   &\ -2.145974046054417415799~\cite{YP:2010}& &            \\
3  &\ -5.04087674559544~\cite{YP:2010}   &\ -5.0408767455954 &\ -5.0408767455954 \\
4  &\ -9.18487389534832~\cite{YP:2010}   &\ -9.1848738953483 &\ -9.1848738953483 \\
5  &\ -14.57852803288144~\cite{YP:2010}  &\ -14.578528032881 &\ -14.578528032880 \\
6  &\ -21.22201770128897~\cite{YP:2010}  &\ -21.222017701289 &\ -21.222017701289 \\
7  &\ -29.115415715499996~\cite{YP:2010} &\ -29.115415715500 &\ -29.115415715500 \\
8  &\ -38.25875730762858~\cite{YP:2010}  &\ -38.258757307629 &\ -38.258757307631 \\
9  &\ -48.65206163912163~\cite{YP:2010}  &\ -48.652061639122 &\ -48.652061639120 \\
10 &\ -60.29534004623167~\cite{YP:2010}  &\ -60.295340046232 &\ -60.295340046233 \\
20 &\ -245.477552~\cite{GD:1988}         &\ {\bf -245.47755786374} &\ -245.47755786754 \\
30 &\ -555.659457~\cite{GD:1988}         &\ {\bf -555.65946790566} &\ -555.65946792636 \\
40 &\ -990.841287~\cite{GD:1988}         &\ {\bf -990.84130097232} &\ -990.84130101968 \\
50 &\ -1551.02309~\cite{GD:1988}         &\ {\bf -1551.0231031980} &\ -1551.0231032767 \\
\hline\hline
\end{tabular}}
\label{table21SE}
\end{table}

\begin{table}
\caption{Energy $E$ of the excited state $2^3S$ for the Helium isoelectronic sequence. The
2nd column contains the results~\cite{YP:2010} for $Z \in [2,10]$; the
energies obtained via the LMM ({\it all 14 printed digits exact}) are in the 3rd column, 
as for $Z=20-50$ (LMM shown in bold), there are no results to compare with - 
this is {\it the first calculation}.
Data obtained via fit~$E_{\{1,2\}}^{(2^3S)}$~\re{fitP95} shown in the 4th column. 
At $Z \rar \infty$, the exact energy~\re{E2H} $E=E_{1,2}=-\frac{5}{8} Z^2$.}
\scalebox{1}{
\begin{tabular}{c|r|r|r}
\hline\hline
$Z$ &\multicolumn{1}{c|}{$E$~\cite{YP:2010}} &\multicolumn{1}{c|}{Present (LMM)} &
\multicolumn{1}{c}{$E_{\{1,2\}}^{(2^3S)}$~\re{fitP95}} \\
\hline
2 &\ -2.175229378236791305738977\ &\,\,\,-2.1752293782368\ &\  -2.17522937823679\\
3 &\ -5.11072737257074\           &\,\,\,-5.1107273725707\ &\  -5.11072737257075\\
4 &\ -9.29716658977762\           &\,\,\,-9.2971665897776\ &\  -9.29716658977760\\
5 &\ -14.7338973488142\           &\ -14.733897348814\     &\ -14.73389734881442\\
6 &\ -21.42075590230796\          &\ -21.420755902308\     &\ -21.42075590230735\\
7 &\ -29.35768173749104\          &\ -29.357681737491\     &\ -29.35768173749166\\
8 &\ -38.54464732008501\          &\ -38.544647320085\     &\ -38.54464732008537\\
9 &\ -48.98163832951831\          &\ -48.981638329518\     &\ -48.98163832951790\\
10&\ -60.66864658407337\          &\ -60.668646584073\     &\ -60.66864658407272\\
20&&\ {\bf -246.28908843655}\ &\ -246.28908843655 \\
30&&\ {\bf -556.90971555002}\ &\ -556.90971555002 \\
40&&\ {\bf -992.53038652813}\ &\ -992.53038652813 \\
50&&\ {\bf -1553.1510746788}\ &\ -1553.1510746788 \\
\hline\hline
\end{tabular}}
\label{T23S}
\end{table}

\begin{table}
\caption{Energies of the second spin-triplet excited state $3^3S$ for the Helium isoelectronic sequence 
({\it the first calculation}), 
obtained via LMM, 2nd column ({\it all 14 printed digits exact}), and by fit $E_{\{1,2\}}^{(3^3S)}$~\re{fitP95}, 
3rd column. At $Z \rar \infty$, 
the exact energy, see~\re{E2H}, is $E=E_{1,3}=-\frac{5}{9} Z^2$.}
\scalebox{1.0}{
\begin{tabular}{c|c|c}
\hline
$Z$& \multicolumn{1}{c|}{Present (LMM)}&\multicolumn{1}{c}{$E_{\{1,2\}}^{(3^3S)}$~\re{fitP95}}  \\
\hline
 2& -2.0686890674724& -2.0686890674724 \\  
 3& -4.7520764560456& -4.7520764560456 \\    
 4& -8.5469720690618& -8.5469720690618 \\    
 5& -13.453104279847& -13.453104279847 \\  
 6& -19.470403018224& -19.470403018224 \\ 
 7& -26.598842151894& -26.598842151894 \\ 
 8& -34.838409733670& -34.838409733670 \\
 9& -44.189099532228& -44.189099532228 \\
10& -54.650907980915& -54.650907980915 \\
20& -220.38026085466& -220.38026085466 \\  
30& -497.22080601297& -497.22080601297 \\  
40& -885.17248151502& -885.17248151502 \\   
50& -1384.2352756587& -1384.2352756587 \\
\hline\hline
\end{tabular}}
\label{tableE33S}
\end{table}

\section{Second critical charge and Puiseux expansion.}

It is evident - on physics grounds - that for any chosen (ground or excited) state of the Helium sequence 
there exists a critical charge $Z=Z_B$ which separates the domain where the Schr\"odinger equation has a
square-integrable, normalizable eigenfunction from the domain where
it does {\it not}. We call this charge $Z_B$ {\it the second critical charge} in order to distinguish it from 
the (first) critical charge $Z_c$ where the energy level enters the continuum, where the ionization energy 
vanishes. The critical charge $Z_B$ can depend on the state under consideration.
It is natural to guess the existence of a singularity (a type of branch point) at $Z=Z_B$ which {\it may} 
define the radius of convergence of the $1/Z$-expansion (see below). This singularity may be related to the level 
crossing. 
This guess was checked (and confirmed) for the ground state $1^1\,S$ energy, see e.g. \cite{AoP:2019} 
and references therein. It turned out that with high accuracy $Z^{(1^1\,S)}_B=0.90485$4, see \cite{TLO:2016}.
It was also shown that for $Z=Z^{(1^1\,S)}_B$ there exists a Puiseux expansion for the energy 
of the ground state $1^1\,S$ having terms with integer and half-integer powers {\it only},
\begin{equation}
\label{PuiseuxGen}
\begin{split}
 E(Z) =& E_{B} + q_{{1}} \left( Z- Z_{B}\right)^{1/2} + p_1 \left( Z-Z_{B} \right)
 + q_{{3}} \left( Z- Z_{B}\right)^{3/2} + p_{{2}} \left( Z - {\it Z_B} \right)^{2}
 + q_{{5}} \left( Z- Z_{B} \right)^{5/2}
\\&
+ p_{{3}} \left( Z- Z_{B} \right)^{3}+q_{{7}} \left(Z - Z_{B} \right)^{7/2}
+ p_{{4}} \left( Z- Z_{B} \right)^{4} + \ldots \ ,
\end{split}
\end{equation}
where $E_{B}=E(Z=Z_{B})$ and $ p_i, q_i, \ i=1,2,3,\ldots$ are numerical coefficients. 
Note that the term of power 1/2 can be absent in this expansion - in folklore this result 
is assigned to Julian Schwinger: the linear term in the expansion comes next after the constant term, 
we confirmed this result. 
In this case, the singularity at $Z=Z_B^{(1^1\,S)}$ is a square-root branch point with the exponent 3/2. 
Coefficients $\{ p, q\}$ are found by making an interpolation~\re{PuiseuxGen} with accuracy 3-5 s.d. of the thirteen energies at 
$Z = 0.904854, 0.91, 0.91028...,0.92,0.93,0.94,0.95,0.96,0.97,0.98,0.99,1.00,2.00$, respectively, 
calculated using the LMM, see \cite{AoP:2019} \footnote{This expansion was never published in complete form}.

Recently, it was checked that an expansion similar to (\ref{PuiseuxGen}) occurs for the energy of the first excited, spin-singlet state $2^1\,S$ at the {\it same} second critical charge, $Z=Z_B^{(2^1\,S)}=Z_B^{(1^1\,S)}=0.904854$, 
see \cite{TLO:2016,Part4}. It was seen that with high accuracy the coefficients $E_{B}=E(Z_{B})$ and $\{ p\}$ in 
(\ref{PuiseuxGen}) remain the same for both states, while the coefficients $\{ q\}$ change sign. 
This Puiseux expansion provides an accuracy of 2-3 s.d.
in the energies for $Z=2,3,\ldots, 6$. This indicates the level crossing of the states $1^1\,S$ and $2^1\,S$ at $Z=Z_B$. 
Thus, these states are analytically-connected. In explicit form both Puiseux expansions have the form
\begingroup
\small
\begin{equation}
\label{1s-2s-spin-singlet}
\begin{split}
E^{{spin-singlet}}_{1^1\,S/2^1\,S} = &
E_B - 1.12349 (Z-Z_{B})
 -/+ 0.19767 (Z-Z_{B})^{3/2}
-0.75285 (Z-Z_{B})^2
\\ &
 -/+0.10828 (Z-Z_{B})^{5/2}
 -0.016 (Z-Z_{B})^3
 +/-0.011 (Z-Z_{B})^{7/2}\ , 
\end{split}
\end{equation}
\endgroup
where the first sign in front of terms of fractional degrees corresponds to the ground state $1^1\,S$, 
while the second sign corresponds to the excited state $2^1\,S$.
Eventually, $Z_B=0.904854$ and $E_B=-0.407932$. 

Technically, in order to find the coefficients $\{ p, q\}$ in~\re{1s-2s-spin-singlet} one can consider the sum $E_+$ and the
difference $E_-$ of the energies of the $1^1\,S, 2^1\,S$ states, 
\[
  E_+ = E^{(1^1\,S)} +  E^{(2^1\,S)}\quad ,\quad E_- = E^{(1^1\,S)} -  E^{(2^1\,S)}\ ,
\]
which are then interpolated in the domain $Z < 6$, see above, via Puiseux expansion. 
We saw that $E_+$ has terms of half-integer degrees with negligibly small coefficients, 
while $E_-$ has terms of integer degrees with negligibly small coefficients. 
Hence, one can draw the conclusion that the expansion of $E_+$ has terms of integer degrees only, 
while the expansion of $E_-$ has terms of half-integer degrees only.
 

In \cite{Part5} it is shown that by making a detailed analysis of the exponential decay rate 
of the accurate variational trial wavefunctions for various $Z$ at large distances 
the second critical charges for the first two spin-triplet states $2^3\,S$ and $3^3\,S$ 
can be found: with high accuracy these critical charges coincide! 
They are equal to $Z_B^{(2^3\,S)}=Z_B^{(3^3\,S)}=0.8799$. By fitting the energies of the
$2^3\,S$ and $3^3\,S$ states at $Z=2-8$ , see Tables \ref{T23S}-\ref{tableE33S}, using a Puiseux expansion 
(\ref{PuiseuxGen}), and requiring an accuracy of 4-6 s.d., we arrive at the following 
Puiseux expansions,
\begingroup
\small
\begin{equation}
\label{2s-3s-spin-triplet}
 \begin{split}
 E^{{spin-triplet}}_{2^3S/3^3S} & = E_B +/-0.011065 (Z-Z_{B})^{1/2}
 -0.884233 (Z-Z_{B}) -/+0.044262 (Z-Z_{B})^{3/2}
\\ &
-0.596588 (Z-Z_{B})^2 -/+0.009601 (Z-Z_{B})^{5/2} +0.001215 (Z-Z_{B})^3
\\ &
 +/-0.000270 (Z-Z_{B})^{7/2} - 0.000082 (Z-Z_{B})^4\ ,
 \end{split}
\end{equation}
\endgroup
where the first sign in front of terms of fractional degrees corresponds to the spin-triplet 
ground state $2^3\,S$, while the second sign corresponds to the spin-triplet excited state $3^3\,S$.
Eventually, $Z^{{spin-triplet}}_B=0.8799$ and $E^{{spin-triplet}}_B=-0.384708$. 
It must be emphasized that in this case the square-root term is present 
in the both Puiseux expansions (\ref{2s-3s-spin-triplet}). Thus, both Puiseux expansions are 
analytically connected indicating the existence of a square-root branch point at $Z=Z_B$ and a
level crossing of the $2^3\,S$ and $3^3\,S$ states.
Let us note that such a value of the second critical charge, $Z_B < 1$, indicates the existence 
of a square-integrable wavefunction for $Z=1$. Since this wavefunction is very much spacially extended 
we were unable to find it. Neither of these two states has been observed experimentally.

\section{$1/Z$-expansion.}

In \cite{Hylleraas} it was shown that for any given eigenstate a perturbation theory 
in $1/Z$ can be developed, see the Hamiltonian (\ref{H}), which is of the form
\begin{equation}
\label{PT}
  E\ =\ \epsilon_0 Z^2 + \epsilon_1 Z + \epsilon_2 + \epsilon_3 /Z + \ldots \ ,
\end{equation}
this is the celebrated $1/Z$-expansion by Hylleraas. 
Here $\epsilon_0=E_{n_1,n_2}$, see (\ref{E2H}), and $\epsilon_1$ is a known rational number 
for all the so far studied eigenstates, some of these rational numbers were explored in the past, 
see e.g. \cite{ALM:1970}, in particular,
\[
  \epsilon_1=\frac{5}{8}\ (1^1\,S)\ ,\ \frac{169}{729}\ (2^1\,S)\ ,\ \frac{137}{729}\ (2^3\,S)\ ,
  \ \frac{3071}{32768}\ (3^3\,S)\ .
\]
Several other perturbative $\epsilon_n$-coefficients are known numerically for some states, 
their concrete values are not important for the present study except for $\epsilon_2$, see below. 
In general, it is important to know the functional form of expansion (\ref{PT}), its first 
two coefficients $\epsilon_{0,1}$ only and sometimes the 3rd coefficient $\epsilon_{2}$ 
(called the electronic correlation energy).
Thanks to Tochiro Kato it is known since 1950es that this perturbation theory has a finite radius 
of convergence. Concrete values for the radius of convergence (for different states) 
have not been settled so far.   

\section{Interpolation.}

Let us introduce a new variable,
\begin{equation}
\label{la}
{\la}^{2}=Z-{Z_B}\ .
\end{equation}
It can be easily verified that by using the variable $\la$ the Puiseux expansion (\ref{PuiseuxGen}) 
becomes a Taylor expansion (at $\la=0$), while the $1/Z$-expansion (\ref{PT}) is transformed 
into a Laurent expansion in $1/\la^2$ with a fourth order pole at $\la=\infty$. It is easy 
to find that the simplest interpolation formula matching these two expansions 
is given by a meromorphic function in $\la$,
\begin{equation}
\label{Int}
  -\,E_{N+4,N}(\la(Z))\ =\ \frac{P_{N+4}(\la)}{Q_N(\la)}\ \equiv\ 
  \mbox{gPade}(N+4/N)_{n_0, n_{\infty}} (\la) \ ,
\end{equation}
which we recognize as a {\it (generalized) two-point Pade approximant}. 
Here $P, Q$ are polynomials
\[
   P_{N+4}=\sum_{k=0}^{N+4} a_k \la^k\ ,\ Q_N=\sum_{k=0}^N b_k \la^k\ ,
\]
of degrees $(N+4)$ and $N$, respectively, where the normalization $Q(0)=1$ is chosen, 
thus, $b_0=1$. In this case the total number of free parameters in (\ref{Int}) is $(2N+5)$. 
It is clear that by choosing $P(0)=E_B$, thus, fixing $a_0 = E_B$, reduces the number of free parameters 
to $(2N+4)$. The interpolation is made in two steps: (i) similarly to the standard Pade approximation 
theory some coefficients in (\ref{Int}) are constrained to reproduce exactly a certain number 
of terms $(n_0)$ in the Taylor expansion at small $\la$ \footnote{It has to be remembered 
that the coefficient in front of $\la^0$ is already fixed equal to $E_B$, hence, 
effectively, $n_0$ is reduced by one. } and also a certain number of terms 
$(n_{\infty})$ in the large $\la$-expansion \footnote{It implies that, in particular, 
some coefficients in front of odd powers of $\la$ are constrained to vanish.}, 
(ii) the remaining undefined {$(2N+5-n_0-n_{\infty})$} coefficients are found 
by fitting the numerical data, 
which is considered to be reliable, by minimizing the $\chi^2$. It is a state-of-the-art 
procedure to choose the optimal $(n_0)$ and $(n_{\infty})$, and then to find concrete values 
for the  remaining free parameters.

After several attempts, mainly involving the ground state, we chose $N=5$, 
which the minimal value of $N$ that leads to correct 13-14 s.d. for the ground 
state energies, see Table~\ref{EnSS21}. Usually, we reproduced {\it exactly} the first four terms in the Laurent expansion (\ref{PT}), 
thus, setting $n_{\infty}=4$, and the first three terms in the Puiseux expansion 
(\ref{PuiseuxGen}), $n_{0}=3$ (where $a_0 = E_B$ already). 
Thus, we consider the (generalized) two-point Pade approximant $\mbox{gPade}(9/5) (\la(Z))_{3,4}$.
The remaining nine free parameters in
\begin{equation}
\label{gPade95}
    \mbox{gPade}(9/5) (\la)_{3,4}\ =\ \frac{E_B+a_1 \la+a_2 \la^2+a_3 \la^3+a_4 \la^4+a_5 \la^5+
    a_6 \la^6+a_7 \la^7+a_8 \la^8+a_9 \la^9}{1+b_1 \la+b_2 \la^2+b_3 \la^3+b_4 \la^4+b_5 \la^5}\ ,
\end{equation}
are found by fitting the available numerical data for the ground state, see \cite{AoP:2019}. 
Data for the excited states from Tables \ref{table21SE}-\ref{tableE33S}, obtained via the LMM, 
can be fitted with the intention of
reproducing 13-14 s.d. in energies correctly. Sometimes, achieving the desired accuracy 
forced us relaxing some constraints, thus, reducing the requirements: $n_{\infty}<4$ and $n_{0}<3$. 
Not always we are able to reach the goal keeping $N=5$. It seems desirable to increase $N$ 
by setting $N > 5$. This possibility was not studied in this article.

With $N=5$ the specified two-point Pad\'e approximant (\ref{gPade95}) we are going to use 
for all four states is given by
\begin{equation}
\label{fitP95}
E_{\{2,2\}}(\la) = \frac{E_B + \al_1\la+\sum_{i=2}^{5}a_i\,\la^i +\al_6 \la^6+\al_7 \la^7+
 \alpha_8 \la^8+a_9\,\epsilon_0 \la^9}{1+b_1 \la+ b_2 \la^2 +b_3 \la^3+b_4 \la^4 +  a_9\,\la^5} \,,
\end{equation}
thus, keeping $a_0=E_B$, with constraints
\begin{eqnarray}
\label{prmstripltS}
\al_1 & - & b_1 E_B = q_1\ , \\
\al_6 &=& \epsilon_0\,(b_2+2\,b_4\,Z_B )+\epsilon_1\,b_4\ (\mbox{implying that}\ 0\,\cdot\la\ \mbox{term}) \ ,\non
\\
\al_7 &=& \epsilon_0\,(b_3+2\,a_9\,Z_B )+\epsilon_1\,a_9\ ,\non \\
\al_8 &=& \epsilon_0\,b_4\ (\mbox{implying that}\ 0\,\cdot\la^3\ \mbox{term}) ,
\non
\end{eqnarray}
where we defined in (\ref{gPade95}) that,
\begin{equation}
\label{normalization}
  a_9 \equiv a_9\,\epsilon_0\ ,\ b_5=a_9\ ,
\end{equation}
which allow us to reproduce exactly the leading term $\epsilon_0 Z^2$ in (\ref{PT}). Eventually, the expression 
(\ref{fitP95}) depends on 9 free parameters with four constraints (\ref{prmstripltS}).

For values of $Z$ close to $Z_B$, expansion of the expression ~\re{fitP95} reproduces 
the term $E_B$ and makes the term of order $(Z-Z_B)^{1/2}$ absent in expansion~\re{1s-2s-spin-singlet}, $q_1=0$ 
(for the ground state) while for large values of $Z$, the coefficients in front 
of $Z^2$ and $Z$~in~\re{PT}~\footnote{equivalently, in front of $\la^4$ and $\la^2$ } 
are reproduced  exactly.

Relaxing the constraints on $\al_1$ (see the 1st line in~\re{prmstripltS}, it relaxes the requirement that
$q_1$ must come from the expansions (7)-(8)) and $\al_6$ (see the 2nd line in~\re{prmstripltS}, 
this removes the constraint on disappearance the term $Z^{1/2}$ (equivalently, $\la$ ) 
in $1/Z$-expansion \re{PT}, which arises due to the change of variable~(\ref{la}), results in the approximation 
\begin{equation}
\label{PA13}
   E_{\{1,2\}}(\la)\ ,
\end{equation}
which provides a more flexible, thus, less restricted P\'ade approximant which reproduces the term 
$E_B$~\re{2s-3s-spin-triplet} for $Z$ close to $Z_B$ exactly and 
the coefficients in front of $Z^2$ and $Z$ for large values of $Z$~\re{PT}.

\subsection{$1^1 S$}

For the spin-singlet ground state ($1s^2$)\,$1^1S$  the coefficients 
in the expansions~\re{1s-2s-spin-singlet} and~\re{PT} are defined by
\begin{center}
\begin{tabular}{lcl}
$Z_B = \,\,\,\,\,0.9048539992$\ , &\hspace{0.8cm}\phantom{.}& $\epsilon_0=-1$\ ,\\
$E_B = -0.407932489$\ ,\ $q_1=0$\ ,  &                         & $\epsilon_1=5/8$\ ,\\
\end{tabular}
\end{center}
The four terms are going to be reproduced in~\re{fitP95} via expansions at 
$Z$ near $Z_B$ (two terms)  and large $Z$ (two terms). 
The next terms in the expansions with coefficients $p_1= - 1.12349$ and $\epsilon_2=-0.15766643$, 
respectively, are reproduced accurately, as shown in Table~\ref{tableHe12sprms}.
Note that the 9 parameters $\{a_2,\dots,a_5,a_9,b_1,\dots,b_4\}$ of the Pad\'e approximant 
$E^{(1^1S)}_{\{2,2\}}$~\re{fitP95} are fixed by fitting the LMM-based data in Table~\ref{EnSS21}.
Although this state was studied extensively in~\cite{AoP:2019}, in the present work 
a new set of parameters is found which allow us to calculate the ground state energies {\it vs.} $Z$ 
more accurately. These parameters are presented in Table~\ref{tableHe12sprms} (left column). 
Note that the simple poles in $\la$, see Table~\ref{tableHe12sprms} (left column), are situated sufficiently far away 
from the physics semi-axis $\la \in [0,\infty)$: this does not lead to significant, visible 
non-monotonicity (waves) of the energy $E^{(1^1S)}_{\{2,2\}}$ {\it vs.} $Z$.
A comparison of the numerical LMM data against the two-point Pad\'e 
approximant ones (see Table~\ref{EnSS21}) shows that the Pad\'e approximant reproduces 
at least 12 s.d. in the energy for all $Z \in [1,50]$. For larger $Z$ the accuracy grows 
becoming {\it absolute} at $Z \rar \infty$.

In conclusion, it must be emphasized that the expression $E^{(1^1S)}_{\{2,2\}}$~\re{fitP95} reproduces 
the term $E_B$ and leaves $(Z-Z_B)^{1/2}$ absent from the expansion~\re{1s-2s-spin-singlet}, 
while the coefficients in front of $Z^2$ and $Z$~in \re{PT} are reproduced exactly.

\begin{table}[!thb]
\caption{Parameters for the spin-singlet states $1^1S$  and $2^1S$~in \re{fitP95}. 
It is also shown: ($i$) the coefficient $p_1$ in front of $(Z-Z_b)$ term in~\re{1s-2s-spin-singlet} 
and $\epsilon_2$
($ii$) the position $\la_i$ and the absolute value $|\la_i|$ of the poles at $i=1,2,3,4,5$
in the Pad\'e approximant (13).}
\begin{center}
\scalebox{1}{
\begin{tabular}{c | r | r }
\hline\hline
& \multicolumn{1}{c|}{$1^1S$} &   \multicolumn{1}{|c}{$2^1S$}\\
\hline
$a_2$&  -5.6981404503284&    6.322467305554\\ 
$a_3$&  -17.989697949178&  -26.598886009815\\
$a_4$&  -18.992880737496&   20.011169176394\\
$a_5$&  -47.232296202503&  -43.386785729570\\
$a_9$&   23.798877891853&   13.165328519353\\
$b_1$&   6.2140006981088&   15.027759860647\\
$b_2$&   11.227896128292&  -17.564125164518\\
$b_3$&   26.368938394814&   31.706075014771\\
$b_4$&   11.41686351435 &  -11.624664168958\\
\hline			                       
 $p_1$  & -1.118  &  -0.843     \\
 {$\epsilon_2$} &  {-0.15766645} & {-0.11451199}\\
\hline
$\lambda_1$    &  $-0.201$          & $-0.062$\\
$\lambda_{2,3}$&  $-0.168\pm i0.477$& $0.040\pm i1.284$\\
$\lambda_{4,5}$&  $ 0.029\pm i0.903$& $0.432\pm i0.749$\\
$|\lambda_1|$    & $0.201$& $0.062$\\
$|\lambda_{2,3}|$& $0.506$& $1.284$\\
$|\lambda_{4,5}|$& $0.903$& $0.864$\\
\hline\hline
\end{tabular}}
\end{center}
\label{tableHe12sprms}
\end{table}

\subsection{$2^1 S$}

Let us now consider the first spin-singlet excited  state ($1s2s$) $2^1S$. In this case, the coefficients 
for expansions~\re{1s-2s-spin-singlet} and~\re{PT} are given by
\begin{center}
\begin{tabular}{lcl}
$Z_B = \,\,\,\,\,0.9048539992$\ ,  & \hspace{0.8cm}\phantom{.}& $\epsilon_0=-5/8$\ ,\\
$E_B = -0.407932489$\ , &           & $\epsilon_1=169/729$\ ,\\
\end{tabular}
\end{center}
they are going to be reproduced in~\re{fitP95} via expansions at $Z\sim Z_b$ and  large Z.
As shown in Table~\ref{tableHe12sprms}, the coefficients $p_1=- 1.12349$ and $\epsilon_2=-0.11451014$ 
in the next terms of expansions are correctly reproduced.
For this state, the values of the 9 parameters $\{a_2,\dots,a_5,a_9,b_1,\dots,b_4\}$ 
of the Pad\'e approximant 
$E^{(2^1S)}_{\{2,2\}}$~\re{fitP95}, obtained by fitting the LMM data, 
are shown in Table~\ref{tableHe12sprms}(right column). 
Note that the simple poles in $\la$, see Table~\ref{tableHe12sprms} (right column), 
are situated sufficiently far away 
from the physics semi-axis $\la \in [0,\infty)$: this does not lead to significant, 
visible non-monotonicity (waves) of the energy $E^{(2^1S)}_{\{2,2\}}$ {\it vs.} $Z$.
A comparison of the numerical results obtained via LMM and those obtained by using the two-point 
Pad\'e approximant (13) is presented in Table~\ref{table21SE}. As can be seen not less than 12 s.d. 
are reproduced for $Z \leq 10$, then the accuracy starts to decrease reaching $9$ s.d. for $Z=40$ 
and then it increases again reaching the absolute accuracy at $Z \rar \infty$. 

Note that $E^{(2^1S)}_{\{2,2\}}$~\re{fitP95} reproduces the first term $E_B$ and leaves out 
the second term 

\noindent
$(Z-Z_B)^{1/2}$ in expansion~\re{1s-2s-spin-singlet}, and reproduces the two terms 
of orders $Z^2$ and $Z$ in~\re{PT}.

\subsection{$2^3 S$}

Now let us consider the lowest energy state in which the electron spins are parallel, thus, 
the total spin equals one:
the spin-triplet state $(1s2s)\,2^3S$. In this case the parameters of expansions found 
in~\cite{ALM:1970,KS:1963}
are given by
\begin{center}
\begin{tabular}{lcl}
$Z_B = \,\,\,\,\,0.87989$\ ,  &\hspace{0.8cm}\phantom{.}& $\epsilon_0=-5/8$\ ,\\
$E_B =-0.3847081$\ ,     &       & $\epsilon_1=137/729$\ ,\\
\end{tabular}
\end{center}
They are going to be reproduced in~\re{fitP95} exactly while making expansions 
at $Z$ around to $Z_B$ (one term) and large $Z$ (two terms), respectively.
The coefficients $q_1\,=\, 0.011065$, $p_1\,=\,-0.884233$  and $\epsilon_2=-0.047409192$ 
in the next terms of the expansions \re{2s-3s-spin-triplet} and \re{PT}, respectively, 
are presented in Table~\ref{tableHe23sprms}.

After fitting the numerical data obtained via LMM the values of the free parameters are
displayed in the 2nd column of  Table~\ref{tableHe23sprms} (left column). 
Note that the simple poles in $\la$, see Table~\ref{tableHe23sprms}  (left column), are situated sufficiently far away 
from the physics semi-axis $\la \in [0,\infty)$: this does not lead to significant, 
visible non-monotonicity (waves) of the energy $E^{(2^3S)}_{\{1,2\}}$ {\it vs.} $Z$.
The expansion of the Pad\'e 
approximant $E^{(2^3S)}_{\{1,2\}}$ (13), see~\re{PA13}  reproduces the value $E_B$ for $Z$ close to 
$Z_B$, see~\re{2s-3s-spin-triplet} and the coefficients in front of $Z^2$ and $Z$ for large 
values of $Z$, see~\re{PT}.
A comparison of the numerically obtained energy values in LMM against the two-point
Padé approximant $E^{(2^3S)}_{\{1,2\}}$ shows that at least 13 s.d. are reproduced
(see Table~\ref{T23S}).

\begin{table}[!thb]
\caption{Parameters for the spin-triplet states $2^3S$  and $3^3S$ in (13), see~\re{PA13}. 
It is also shown: (i) the coefficient $q_1$ in front of  $(Z-Z_b)^{1/2}$ in~\re{2s-3s-spin-triplet}
(ii) the position $\la_j$ and norm $|\la_j|$ of the five poles $j=1,2,3,4,5$ in the two-point 
Pad\'e approximant (13).}
\begin{center}
\scalebox{1}{
\begin{tabular}{c | r | r }
\hline\hline
& \multicolumn{1}{c|}{$2^3S$} &   \multicolumn{1}{|c}{$3^3S$}\\
\hline
$a_1$&   0.1346410928570494& 0.7062083507800702\\
$a_2$&  -2.9863549938707585& -2.743811129460274\\ 
$a_3$&  -1.1579741129463628&  1.591050751980635\\
$a_4$&   -5.585962716446034& -4.449017775986878\\
$a_5$&   -8.312824729944447& -2.606064850185295\\
$a_6$&  -3.0022870605168266& -1.697461065068710\\
$a_9$&    13.45861688201462&  9.382210505608663\\
$b_1$&  -0.4044825030247633& -1.858775038114970\\
$b_2$&    5.681523880367753&  4.904201803147925\\
$b_3$&   3.9922714188898416&  0.083447946191657\\
$b_4$&  -0.6016502817273021& -1.161956671468978\\
\hline			                       
 $q_1$                      &  -0.0210        &  -0.009  \\
 $p_1$  & -0.8091         & -0.8736\\
 $\epsilon_2$&-0.0474  &-0.0323\\
\hline
$\lambda_1$    &  $-0.719$           & $-0.912$           \\
$\lambda_{2,3}$&  $  0.063\pm i0.414$& $  0.202\pm i0.489$\\
$\lambda_{4,5}$&  $  0.319\pm i0.699$& $  0.315\pm i0.563$\\
$|\lambda_1|$    & $0.719$& $0.912$\\
$|\lambda_{2,3}|$& $0.418$& $0.530$\\
$|\lambda_{4,5}|$& $0.769$& $0.646$\\
\hline\hline
\end{tabular}}
\end{center}
\label{tableHe23sprms}
\end{table}

\subsection{$3^3 S$}
For the second spin-triplet excited state ($1s3s)\,3^3S$ the parameters found in~\cite{ALM:1970} are

\begin{center}
\begin{tabular}{lcl}
$Z_B = \,\,\,\,\,0.87989$\ ,   & \hspace{0.8cm}\phantom{.}& $\epsilon_0=-5/9$\ ,\\
$E_B =-0.3847081$\ ,           &  & $\epsilon_1=3071/32768$\ .\\
\end{tabular}
\end{center}
They are going to be reproduced in~\re{fitP95} via expansions at $Z\sim Z_b$  and large $Z$. 
The $q_1\,=\, -0.011065$, $p_1=-0.884233$  and $\epsilon_2=-0.032158$ are coefficients of the 
expansion \re{2s-3s-spin-triplet} and \re{PT}, respectively, obtained from the fit are shown 
in Table~\ref{tableHe23sprms}.
After fitting the numerical data obtained via LMM the values of the free parameters are
displayed in the third column of  Table~\ref{tableHe23sprms} (right column). 
Note that the simple poles in $\la$, see Table~\ref{tableHe23sprms}  (right column), 
are situated sufficiently far away 
from the physics semi-axis $\la \in [0,\infty)$: this does not lead to significant, 
visible non-monotonicity (waves) of the energy $E^{(3^3S)}_{\{1,2\}}$ {\it vs.} $Z$. 
The Pad\'e approximant $E^{(3^3S)}_{\{1,2\}}$~\re{PA13}, see (13)
reproduces one and the two terms in the expansions at small $\la$ and large $Z$  
in~\re{2s-3s-spin-triplet} and ~\re{PT}, respectively.

Table~\ref{tableE33S} shows a comparison of the energy values obtained using the LMM and those
from the Pad\'e approximant $E_{\{1,2\}}^{(3^3S)}$~\re{fitP95} for $Z \leq 50$.
Surprisingly, all 14 s.d. coincide! 

\section{Conclusions}

In this article we showed that discretizing three-dimensional space by introducing a
non-uniform lattice with nodes related to the zeroes of the Laguerre polynomials by using the LMM allows us
to solve the Schr\"odinger equation for the $S$ states of the helium-like sequence numerically 
with extremely high accuracies. 
For the four low-lying states of the helium-isoelectron sequence $ 1^1S, 2^1S, 2^3S, 3^3S $ 
the lattice of modest size $50 \times 50 \times 40$ leads to an accuracy of 14-15 s.d. in the energies 
for nuclear charges $Z \leq 50$. In many cases the highly accurate calculations were carried out 
for the first time. It seems evident that 
for the higher excited $S$ states similar accuracies can be reached with the same lattice size.
By increasing the lattice size one can achieve even higher accuracies~\cite{HB:1999,HB:2001}: 
this was checked by making 
a detailed analysis of the one-dimensional quartic anharmonic oscillator and the quartic double-well potential 
\cite{TV:2021}. The LMM provides the opportunity to study wave functions, thus, to calculate polarizabilities, 
transition amplitudes etc, which will be done elsewhere. For the $P$ states of the helium-like sequence 
the LMM calculations will require using the six-dimensional Schr\"odinger equation as well 
as the $S$ states of lithium-like sequence. This will be done elsewhere.

In this paper an interpolation formula of general nature is introduced, it is applied to the energies 
of any excited state of the Helium-like sequence and, in principle, of any atomic system. 
It is based on matching the $1/Z$-expansion at large $Z$ and the Puiseux expansion 
with terms of integer and half-integer degrees around the so-called second critical charge $Z_B$, 
hinted by F and D Stillinger (1969, 1974) \cite{Stillinger:1966}, 
confirmed by the present authors in 2019 \cite{AoP:2019} for the ground state $1^1 S$ 
and extended to the excited states in the present work. It has the form of a two-point Pade approximant 
$\mbox{gPade}(N+4/N)(\la)=P_{N+4}(\la)/Q_N(\la)$ of order $N$ as the ratio of two polynomials 
in the argument $\la \equiv (Z-Z_B)^{1/2}$. 
As an example, for the first two spin-singlet $1^1 S$, $2^1 S$ and 
the first two spin-triplet $2^3 S$, $3^3 S$ states this interpolation formula of the order five, $N=5$, 
is presented explicitly with specific parameters. It allows us to achieve the accuracy of 10-14 s.d. 
in the energies for any physically-relevant nuclear charge $Z$ of the helium-like sequence, while 
in the limit $Z \rar \infty$ the accuracy becomes absolute.

The same interpolation formula (with different parameters and $Z_B=2.0090$) was applied to the ground state energy 
of the Lithium-like sequence, see \cite{AoP:2019}: it was checked that the accuracy of 13 s.d. was 
achieved for the nuclear charge $Z \leq 20$, while in the limit $Z \rar \infty$ the accuracy becomes 
absolute. The case of excited states will be considered elsewhere.

A $1/Z$-expansion of the same functional form can be constructed for {\it any} atomic system with 
fixed number of electrons in any arbitrary excited state. From a physics viewpoint it is evident that 
the second critical charge $Z_B$ must exist, and that it should be around $(k-1)$, 
where $k$ is the number of electrons. It is natural to guess that the Puiseux expansion 
with terms of integer and half-integer degrees should exist as well.
This allows us to construct the interpolation $\mbox{gPade}(N+4/N)(\la)$ and the finding of free parameters
including $Z_B$ by fitting experimental or numerical data.
 
The existence of mass corrections of the order of $m_e/M_{nuclei}$, QED corrections $\sim \al^2$ and 
relativistic corrections of $(v/c)^2$ reduces the domain of applicability of non-relativistic theory 
of helium-sequence to four s.d. in total energy, see e.g. \cite{AoP:2019}. It is worth noting 
that at large $Z \sim 100$ the relativistic corrections become significant: 
the Schr\"odinger equation must be replaced by the Dirac one, there exists a critical charge 
$Z_{critical} \sim 137$, vacuum is rearranged and physics becomes completely different 
at $Z > 137$, see \cite{QED:137}. Straightforward estimate shows that two-point Pade approximant 
$\mbox{gPade}(5/1)_{1,2}(\la)$ with five free parameters describes 3 d.d. of the energies of 
$1^1 S$ state for any $Z \in [1, 50]$, thus, the domain of applicability of the non-relativistic 
static approximation. 

\section*{Acknowledgments}

This work is supported in part by the PAPIIT grant {\bf IN104125} (Mexico).


\begin{thebibliography}{99}

\bibitem{Hylleraas}
        E.A.~Hylleraas, \\
        {\it Neue Berechnung der Energie des Heliums im Grundzustande,
        sowie des tiefsten Terms von Ortho-Helium},\\
        {\em Z.~Phys. \bf 54} 347-366 (1929) (in German);\\
        English translation: Quantum chemistry: classic scientific papers, translated and edited
        by H Hettema,  Singapore; London:  World Scientific, 2000, pp 104-121.

\bibitem{Nakashima:2007}
         H.~Nakashima, H.~Nakatsuji,\\
        {\it Solving the Schr\"odinger equation for helium atom and its isoelectronic
        ions with the free iterative complement interaction (ICI) method,}\\
         {\it J. Chem. Phys. \bf 127}, 224104 (2007)
         
\bibitem{Nakashima:2008}
         H.~Nakashima, H.~Nakatsuji,\\
        {\it Solving the electron-nuclear Schrödinger equation of helium atom 
        and its isoelectronic ions with the free iterative-complement-interaction method,}\\
         {\it J. Chem. Phys. \bf 128}, 154107 (2008)        

\bibitem{Korobov:2018}
          D.T.~Aznabaev, A.K.~Bekbaev, V.I.~Korobov,\\
          \textit{Nonrelativistic energy levels of helium atom},\\
         {\em Phys. Rev. \bf A 98} (2018) 012510

\bibitem{YP:2010} 
         V.A.~Yerokhin, K.~Pachucki,\\
         {\it Theoretical Energies of Low-Lying States of Light Helium-Like Ions},\\
         {\em Phys. Rev. \bf A 81} (2010) 022507           
         
\bibitem{OT:2015} 
         H.~Olivares Pil\'on and A.V.~ Turbiner,\\
         \textit{Nuclear critical charge for two-electron ion at Lagrange
          mesh method},\\
         {\it Physics Letters A \bf 379} (2015) 688-690       
         
\bibitem{HB:2001}
	     M.~Hesse and D.~Baye,\\
         \textit{Lagrange-mesh calculations of excited states of three-body atoms and molecules},\\
         {\em J. Phys. B: At. Mol. Opt. Phys. \bf 34}, 1425–1442 (2001)       
         
\bibitem{Baye:2015}
	     D.~Baye,\\
         \textit{Lagrange-mesh method},\\
         {\em Phys. Repts. \bf 565}, 1 - 107 (2015) 
         
\bibitem{AoP:2019}
	     A.V.~ Turbiner, J.C.~ L\'opez Vieyra and H.~Olivares Pil\'on,\\
	     \textit{Few-electron atomic ions in non-relativistic QED:
         Ground state energy},\\
         {\it Annals of Physics \bf 409} (2019) 167908 (19 pp)                   

\bibitem{Stillinger:1966}
        F.H.~Stillinger,\\
        \textit{Ground energy of two-electron atoms},\\
        {\em J. Chem. Phys. \bf 45}, 3623-3631 (1966);\\
        F.H.~Stillinger, D.K.~Stillinger,\\
        \textit{Non-linear variational study of perturbation theory for atoms and ions},\\
        {\em Phys. Rev. \bf 10}, 1109 (1974)
        
\bibitem{HB:1999}
	     M.~Hesse and D.~Baye,\\
         \textit{Lagrange-mesh calculations of three-body atoms and molecules},\\
         {\em J. Phys. B: At. Mol. Opt. Phys. \bf 32}, 5605–5617 (1999)       
        
\bibitem{TME:2017}
        A.V.~Turbiner, W.~Miller, Jr., and M.A.~Escobar-Ruiz,\\
        \textit{Three-body problem in 3D space: Ground state, (quasi)-exact solvability,}\\
        {\em J. Phys. A: Math. Theor. \bf 50}, 215201 (2017) (19pp)
        
        
\bibitem{GD:1988H}
         G.W.F.~Drake,\\
         \textit{High precision variational calculations for the $1s^2 \phantom{.}^{1} S$ 
         state of H$^-$ and\\ the $1s^2\, \phantom{.}^{1}S$ , $1s2s\, ^1S$ and 
         $1s2s\, ^3S$ states of helium},\\
         {\em Nucl. Instrum. Methods Phys. Res. B: Beam Interact. Mater. At. \bf 31}, 7-13 (1988)
                  
\bibitem{GD:2006}
         G.W.F.~Drake (Editor),\\
         \textit{Springer Handbook of Atomic, Molecular, and Optical Physics},
         {\em Springer}, 2006 (Chapter 11)              

\bibitem{GD:1988}
         G.W.F.~Drake,\\
         \textit{Theoretical energies for the n=1 and 2 states of the helium isoelectronic sequence up to Z=100},\\
         {\em Can. J. Phys. \bf 66}, 586-611 (1988)
                    
\bibitem{TLO:2016}
	     A.V.~Turbiner, J.C.~L\'opez Vieyra and H.~Olivares Pil\'on,\\
	     \textit{Three-body quantum Coulomb problem: Analytic continuation},
	     {\em Mod. Phys. Lett. \bf A 31} (2016) 1650156             
        
\bibitem{Part4}
         J.C.~Lopez Vieyra and A.V.~Turbiner,\\
         \textit{Ultra-Compact accurate wave functions for He-like iso-electronic
         sequences and variational calculus. IV. Spin-singlet states $(1s\,ns)$
         $n\,{}^1 S$ family of the Helium sequence},\\
         {\it Advances in Quantum Chemistry \bf 89} (2023) Chapter 5, 305-339\\
         (to the memory of F.E.~Harris)\\
         Edited by E.J.~Brandas and R.J.~Bartlett 
         
\bibitem{Part5}
         J.C.~Lopez Vieyra and A.V.~Turbiner,\\         
         \textit{Ultra-Compact accurate wave functions for He-like iso-electronic
         sequences and variational calculus. V. Spin-triplet states $(1s\,ns)$
         $n\,{}^3 S$ (with $n>1$) family of the Helium sequence},\\
         (work in progress)             
                  
\bibitem{ALM:1970}  
         K.~Aashamar, G.~Lyslo, and J.~Midtdal, \\
         \textit{Variational Perturbation Theory Study of Some Excited States of Two-Electron Atoms},\\
         {\em J. Chem. Phys. \bf 52}, 3324-3336 (1970)
 
\bibitem{KS:1963}
         R.E.~Knight and C.W.~Scherr,\\
         \textit{Two-Electron Atoms II. A Perturbation Study of Some Excited States},
         {\em Rev. Mod. Phys. \bf 35}, 431-436 (1963)     
      
\bibitem{TV:2021}
         A.V.~Turbiner and J.C.~del Valle,\\
         \textit{Comment on: Uncommonly accurate energies for the general quartic
         oscillator,\\ 
         Int. J. Quantum Chem., e26554 (2020), by P.Okun and K.Burke},\\
         {\it Int. Journal of Quantum Chemistry \bf 122} (2021) qua.26766 (pp.4)\\
         ( early view: July 2021)
         
         
\bibitem{QED:137}         
         Ya.B.~Zeldovich and V.S.~Popov,\\
         \textit{Electronic structure of superheavy atoms},\\
         {\it Uspekhi Fiz. Nauk. \bf 105} (1971) 403-440;\\
         {\it Soviet Physics Uspekhi, \bf 14(6)} (1972) 673–694 (English translation)
          
\end{thebibliography}
\end{document}